\documentclass[11pt]{article}
\usepackage{framed,epsf,latexsym,amsmath,amssymb,amscd,textcomp}
\usepackage[numbers]{natbib}                                 
\usepackage{graphicx}
\usepackage[pdftex,colorlinks]{hyperref}
\usepackage[margin=1.0in]{geometry}

\newtheorem{theorem}{Theorem}[section]
\newtheorem{lemma}[theorem]{Lemma}

\newtheorem{counter-example}[theorem]{Counter example}

\newtheorem{assumption}[theorem]{Assumption}
\newtheorem{open question}[theorem]{Open question}

\newcommand{\ca}{{\cal A}}

\newcommand{\cd}{{\cal D}}

\newcommand{\cg}{{\cal G}}
\newcommand{\ch}{{\cal H}}

\newcommand{\cl}{{\cal L}}

\newcommand{\cx}{{\cal X}}

\DeclareMathOperator*{\sign}{sign}

\DeclareMathOperator*{\maj}{MAJ}

\newcommand{\Uni}{\mathrm{Uni}}
\newcommand{\Fr}{\mathrm{Fr}}

\newcommand{\reals}{{\mathbb R}}

\newcommand{\csp}{\mathrm{CSP}}

\newcommand{\rand}{\mathrm{rand}}

\newcommand{\xor}{\mathrm{XOR}}
\newcommand{\pol}{\mathrm{POL}}
\newcommand{\mxor}{\mathrm{MXOR}}

\newcommand{\val}{\mathrm{VAL}}
\newcommand{\half}{\mathrm{HALF}}

\newcommand{\proof}{{\par\noindent {\bf Proof}\space\space}}
\newcommand{\proofbox}{\hfill $\Box$}

\DeclareMathOperator{\Err}{Err}
\DeclareMathOperator{\poly}{poly}

\newcommand{\inner}[1]{\langle #1 \rangle}

\title{Complexity Theoretic Limitations on Learning Halfspaces}

\author{Amit Daniely\thanks{Department of Mathematics, The Hebrew University and Microsoft Research Herzliya.}
}

\begin{document}
\maketitle
\setcounter{page}{0}

\thispagestyle{empty}
\maketitle

\begin{abstract}
We study the problem of agnostically learning halfspaces which is defined by a fixed but unknown distribution $\cd$ on $\mathbb{Q}^n\times \{\pm 1\}$. We define $\Err_{\half}(\cd)$ as the least error of a halfspace classifier for $\cd$. A learner who can access $\cd$ has to return a hypothesis whose error is small compared to $\Err_{\half}(\cd)$.

Using the recently developed method of \cite{daniely2013average} we prove hardness of learning results assuming that random $K$-$\xor$ formulas are hard to (strongly) refute.
We show that no efficient learning algorithm has non-trivial worst-case performance even under the guarantees that $\Err_{\half}(\cd) \le \eta$ for arbitrarily small constant $\eta>0$, and that $\cd$ is supported in $\{\pm 1\}^n\times \{\pm 1\}$. Namely, even under these favorable conditions, and for every $c>0$, it is hard to return a hypothesis with error $\le \frac{1}{2}-\frac{1}{n^c}$. In particular, no efficient algorithm can achieve a constant approximation ratio.
Under a stronger version of the assumption (where $K$ can be poly-logarithmic in $n$), we can take $\eta =  2^{-\log^{1-\nu}(n)}$ for arbitrarily small $\nu>0$. These results substantially improve on previously known results~\cite{danielySh2014, FeldmanGoKhPo06, KlivansKo2014, KlivansSh06, KalaiKlMaSe05}, that only show hardness of exact learning.
\end{abstract}

\newpage

\section{Introduction}
% Definition
In the problem of agnostically learning halfspaces, a learner is given an access to a distribution $\cd$ on $\mathbb{Q}^n\times \{\pm 1\}$. The goal is to output\footnote{Throughout, we require algorithms to succeed with a constant probability (that can be standardly amplified by repetition).} (a description of) a classifier $h:\mathbb{Q}^n\to \{\pm 1\}$ whose error, $\Err_{\cd}(h):=\Pr_{(x,y)\sim\cd}\left(h(x)\ne y\right)$, is small comparing to $\Err_{\half}(\cd)$ -- the error of the best classifier of the from $h_w(x)=\sign(\inner{w,x})$. We say that a learning algorithm learns halfspaces if, given an accuracy parameter $\epsilon>0$, it outputs a classifier with error at most $\Err_{\half}(\cd)+\epsilon$. The learner is {\em 
efficient} if it runs in time $\poly\left(n,\frac{1}{\epsilon}\right)$ and the output hypothesis can be evaluated in $\poly\left(n,\frac{1}{\epsilon}\right)$ time given its description. The learner has an approximation ratio $\alpha=\alpha(n)$ if it is guaranteed to return $h$ with
$\Err_{\cd}(h)\le \alpha\cdot \Err_{\half}(\cd)+\epsilon$.
We emphasize that we consider the general, {\em improper}, setting where the learner has the freedom to return a hypothesis that is not a halfspace classifier.

% Importance and centrality of learning halfspaces, Poorly understood preior to this work, 
The problem of learning halfspaces is as old as the field of machine learning, starting with
the perceptron algorithm~\cite{Rosenblatt62,Rosenblatt58}, through the modern SVM~\cite{CortesVa95, Vapnik98, ScholkopfSm02, CristianiniSh00, CristianiniSh04} and AdaBoost~\cite{schapire2012boosting, Schapire89, FreundSc95}. Halfspaces are widely used in practice, have been extensively studied theoretically, and in fact motivated much of the existing theory, both the statistical and the computational.

Despite all that, the gap between the performance of best known algorithms and best known lower bounds is dramatic. Best known efficient algorithms \cite{KearnsLi93} for the problem have a poor approximation ratio of $\tilde{\Omega}\left(n\right)$, and have performance better than trivial only when $\Err_{\half}(\cd)\le \tilde{\Theta}\left(\frac{1}{n}\right)$.
As for lower bounds, strong $\mathcal{NP}$-hardness results are known~\cite{arora1993hardness, amaldi1998approximability, GuruswamiRa06, FeldmanGoKhPo06} only for algorithms that are restricted to return a halfspace classifier (a.k.a. proper algorithms). For general algorithms, no $\mathcal{NP}$-hardness results are known, yet several results~\cite{danielySh2014, FeldmanGoKhPo06, KlivansKo2014, KlivansSh06, KalaiKlMaSe05} show that it is hard to agonstically learn halfspaces under several (cryptographic and average case) complexity assumptions. However, these results are quantitatively very weak, as they only rule out exact learning (i.e., with approximation ratio $1$). For example, they do not rule out algorithms that predict only $1.001$ times worst than the best halfspace classifier.

The main result of this paper is a quantitatively strong hardness results, assuming that (strongly) refuting random $K$-$\xor$ formulas is hard. Using the recently developed framework
of the author with Linial and Shalev-Shwartz~\cite{daniely2013average}, we show that for arbitrarily small constant $\eta>0$ and every $c>0$, no $\poly(n)$-time algorithm can return a hypothesis with error $\le \frac{1}{2}-\frac{1}{n^c}$, even when it is guaranteed that $\Err_{\half}(\cd) \le \eta$, and that $\cd$ is supported in $\{\pm1 \}^n\times\{\pm1\}$. This implies in particular, that there is no efficient learning algorithm with a constant approximation ratio.
Under a stronger version of the assumption (where $K$ is allowed to be poly-logarithmic in $n$), we can take $\eta =  2^{-\log^{1-\nu}(n)}$ for arbitrarily small $\nu>0$. This implies hardness of approximation up to a factor of $2^{\log^{1-\nu}(n)}$. Interestingly, this is even stronger than the best known results \cite{arora1993hardness,GuruswamiRa06,FeldmanGoKhPo06} for proper algorithms.

% Results

\subsection{The random $K$-XOR assumption} 
% Why no P \ne NP?
Unless we face a dramatic breakthrough in complexity theory,
it seems unlikely that hardness of learning can be established on standard complexity assumptions such as $\mathbf{P}\ne\mathbf{NP}$ (see \cite{ApplebaumBaXi08,daniely2013average}). Indeed, all currently known lower bounds are based on cryptographic and average-case assumptions. 
One type of such assumptions, on which we rely in this paper, concern the random $K$-$\xor$ problem. This problem has been extensively studied, and assumptions about its intractability were used to prove hardness of approximation results~\cite{Alekhnovich03}, establish public-key cryptography \cite{Alekhnovich03, ApplebaumBaWi10}, and statistical-computational tradeoffs \cite{barakMo2015}.

% Definitions
A {\em $K$-tuple} is a mapping $C:\{\pm 1\}^n\to \{\pm 1\}^K$ in which each output coordinate is a literal and the $K$ literals correspond to $K$ different variables. 
The collection of $K$-tuples is denoted $\cx_{n,K}$. 
A {\em $K$-formula} is a collection $J=\{C_1,\ldots,C_m\}$ of $K$-tuples.
An instance to the {\em $K$-$\xor$ problem} is a $K$-formula, and the goal is to find an assignment $\psi\in \{\pm 1\}^n$ that maximizes
$\val_{\psi,\xor}(J):=\frac{|\{j : \xor_{K}(C_i(\psi))=1\}|}{m}$. We define the {\em value} of $J$ as $\val_\xor(J):=\max_{\psi\in\{\pm 1\}^n}\val_{\psi,\xor}(J)$. We will allow $K$ to vary with $n$ (but still be fixed for every $n$). For example, we can look of the $\lceil\log(n)\rceil$-$\xor$ problem.

% The problem
We will consider the problem of distinguishing random formulas  from formulas with high value.
Concretely, for $m=m(n)$, $K=K(n)$ and $\frac{1}{2}>\eta= \eta (n) > 0$, we say that the problem $\csp^{\rand,1-\eta}_m(\xor_{K})$ is easy, if there exists an efficient randomized algorithm, $\ca$ with the following properties. Its input is a $K$-formula $J$ with $n$ variables and $m$ constraints and its output satisfies:
\begin{itemize}
\item If $\val_{\xor}(J)\ge 1-\eta$, then
\[
\Pr_{\text{coins of }\ca}\left(\ca(J)=\text{``non-random"}\right)\ge\frac{3}{4}
\]  
\item If $J$ is random\footnote{To be precise, the $K$-tuples are chosen uniformly, and independently from one another.}, then with probability $1-o_n(1)$ over the choice of $J$,
\[
\Pr_{\text{coins of }\ca}\left(\ca(J)=\text{``random"}\right)\ge \frac{3}{4}~.
\]  
\end{itemize}
% Best know resuts
It is not hard to see that the problem gets easier as $m$ gets larger, and as $\eta$ gets smaller. For $\eta=0$, the problem is actually easy, as it can be solved using Gaussian elimination. However, even for slightly larger $\eta$'s the problems seems to become hard.
For example, for any constant $\eta>0$, best known algorithms~\cite{feige2004easily,coja2004strong,coja2010efficient, barakMo2015, allenOdWi2015} only work with $m=\Omega\left(n^{\frac{K}{2}}\right)$. In light of that, we put forward the following two assumptions.

% Assumptions
\begin{assumption}\label{hyp:xor_weak}
There are constants $c>0$ and $\frac{1}{2}> \eta>0$ such that for every $K$ and $m=n^{c\log(K)\sqrt{K}}$, the problem $\csp^{\rand,1-\eta}_{m}(\xor_{K})$ is hard.
\end{assumption}

\begin{assumption}\label{hyp:xor_strong}
There are constants $c>0$ and $\frac{1}{2}> \eta>0$ such that for every $s$, $K=\log^s(n)$ and $m=n^{cK}$, the problem $\csp^{\rand,1-\eta}_{m}(\xor_{K})$ is hard.
\end{assumption}

% More evidence
We outline below some evidence to the assumptions, in addition to known algorithms' performance.

{\bf Hardness of approximation.} Hastad's celebrated result~\citep{haastad2001some} asserts that if $\mathbf{P}\ne \mathbf{NP}$, then for every $\eta>0$, it is hard to distinguish
$K$-$\xor$ instances with value $\ge 1-\eta$ from instances with
with value $\le \frac{1}{2}+\eta$. Since the value of a random formula is approximately $\frac{1}{2}$, we can interpret Hastad's result as claiming that it is hard to distinguish formulas with value $1-\eta$ from ``semi-random" $K$-$\xor$ formulas (i.e., formulas whose value is approximately the value of a random formula). Therefore, our assumptions can be seen as a strengthening of Hastad's result.

{\bf Hierarchies and SOS lower bounds.} A family of algorithms whose performance has been analyzed are convex relaxations~\cite{buresh2003rank, schoenebeck2008linear, alekhnovich2005towards} that belong to certain {\em hierarchies} of convex relaxations. Among those hierarchies, the strongest is the Lasserre hierarchy (a.k.a. Sum of Squares). Algorithms from this family achieves state of the art results for the $K$-$\xor$ and many similar problems.
In \cite{schoenebeck2008linear} it is shown that relaxations in the Lasserre hierarchy that work in sub-exponential time cannot solve $\csp^{\rand,1-\eta}_{n^{\frac{K}{2}-\epsilon}}(\xor_{K})$ for any $\eta,\epsilon>0$.

{\bf Lower bounds on statistical algorithms.} Another family of algorithms whose performance has been analyzed are the so-called statistical algorithms.
Similarly to hierarchies lower bounds, the results in~\cite{FeldmanPeVe2015} imply that statistical algorithms cannot solve $\csp^{\rand,1-\eta}_{n^{\frac{K}{2}-\epsilon}}(\xor_{K})$ for any $\eta,\epsilon>0$.

{\bf Resolution lower bounds.} The length of resolution refutations of random $K$-$\xor$ formulas have been extensively studied (e.g.~\cite{haken1985intractability,BeamePi96,BeameKaPiSa98,BenWi99}). It is known (Theorem 2.24 in~\cite{ben2001expansion}) that random formulas with $n^{\frac{K}{2}-\epsilon}$ constraints only have exponentially long resolution refutations. This shows that yet another large family of algorithms (the so-called Davis-Putnam algorithms~\cite{davis1962machine}) cannot efficiently solve $\csp^{\rand,1}_{n^{\frac{K}{2}-\epsilon}}(\xor_{K})$ for any $\epsilon>0$.

{\bf Similar assumptions.} Several papers relied on similar assumptions.
Alekhnovich~\cite{Alekhnovich03} assumed that $\csp^{\rand,1-\eta}_{Cn}(\xor_{3})$
is hard for some $\eta<\frac{1}{2}$ and for every $C>0$. Applebaum, Barak and Wigderson~\cite{ApplebaumBaWi10} assumed that $\csp^{\rand,1-n^{-0.2}}_{n^{1.4}}(\xor_{3})$ is hard. Barak 
and Moitra~\cite{barakMo2015} made the assumption that 
$\csp^{\rand,1-\eta}_{n^{\frac{K}{2}-\epsilon}}(\xor_{K})$ 
is hard for every $\epsilon>0$ and $\eta=\frac{1}{2}-o(1)$. Assumptions on predicates different than $K$-$\xor$ were made in~\cite{daniely2013more,Feige02}. The assumption in~\cite{Feige02} implies that $\csp^{\rand,1-\eta}_{Cn}(\xor_{3})$ is hard for every $C>0$ and $\eta>0$. The assumption in~\cite{daniely2013more} implies that $\csp^{\rand,1-\eta}_{n^{1.5-\epsilon}}(\xor_{3})$ is hard for every $\epsilon>0$ and $\eta>0$. A much more general assumptions was made in~\cite{BarakKS13}. It implies in particular that $\csp^{\rand,1-\eta}_{Cn}(\xor_{K})$ is hard for every $C>0$ and $\eta>0$.

\subsection{Previous Results and Related work}

\paragraph{\bf Upper bounds.} When $\Err_{\half}(\mathcal{D})=0$, the problem of learning halfspaces can be solved efficiently using linear programming. However, even for slightly larger error values the problem seems to become much harder. Currently best known algorithms \cite{KearnsLi93} have non trivial performance only when it is guaranteed that $\Err_{\half}(\mathcal{D})\le \frac{\log(n)}{n}$. This algorithm also achieves approximation ratio of $n$, which is currently the best known approximation ratio. Better guarantees are known under various assumptions on the marginal distribution \cite{AwasthiBalcanLong14, BlaisOdWi08, daniely2015, KalaiKlMaSe05}. For example, a PTAS is known~\cite{daniely2015} when the marginal distribution is uniform.

\paragraph{\bf Lower bounds for general algorithms.} Several hardness assumptions imply that it is hard to agnostically learn halfspaces. Feldman, Gopalan, Khot and Ponnuswami~\cite{FeldmanGoKhPo06} have shown that based on the security of the Ajtai-Dwork cryptosystem. Kalai, Klivans, Mansour and Servedio showed the same conclusion~\cite{KalaiKlMaSe05} based on the hardness of learning parity with noise. Daniely and Shalev-Shwartz~\cite{danielySh2014} derived the same conclusion based on the hardness of refuting random $K$-SAT formulas. Klivans and Kothari~\cite{KlivansKo2014} showed that assuming that learning sparse parity is hard, it is hard to learn halfspaces even when the marginal distribution is Gaussian. We note that all these  results only rule out exact algorithms, but say nothing about approximation algorithms. We note however that by \cite{aspnes1994expressive, beigel1991perceptron}, an algorithm with non trivial performance of $\eta$-realizable distributions will result with quasi-polynomial time algorithm for learning constant depth circuits.

\paragraph{\bf Lower bounds for proper algorithms and hardness of approximation.}
When we restrict the algorithms to return a halfspace classifier, the problem of learning halfspaces is essentially equivalent~\cite{PittVa88} to the computational problem of {\em minimizing disagreements}. In this problem we are given a sample $S=\{(x_1,y_1),\ldots, (x_m,y_m)\}\in \mathbb{Q}^n\times \{\pm 1\}$, and the goal is to find a vector $w\in \mathbb{Q}^n$ that minimizes the fraction of pairs with $\sign(\inner{w,x_i})\ne y_i$. The optimal fraction is called the {\em error} of the sample. 
As a ``standard" and basic computational problem, much is known about it. The problem have been shown $\mathcal{NP}$-hard already in Karp's famous paper \cite{Karp72}. Soon after the discovery of the PCP theorem, Arora, Babai, Stern and Sweedyk~\cite{arora1993hardness} have 
shown that assuming that no quasi-polynomial time algorithm can solve $\mathcal{NP}$-hard problems, there is no efficient algorithm with an approximation ratio of $2^{\log^{1-\epsilon} (n)}$. Later on, the corresponding maximization problem was considered by several authors~\cite{AmaldiKa1998, BenDavidEiLo2003, BshoutyBu2006, GuruswamiRa06, FeldmanGoKhPo06}, culminating with Feldman, Gopalan, Khot and Ponnuswami~\cite{FeldmanGoKhPo06} who showed that assuming that no quasi-polynomial time algorithm can solve $\mathcal{NP}$ hard problems, no efficient algorithm can 
distinguish samples with error $\ge \frac{1}{2}-2^{-\sqrt{\log(n)}}$ from samples with error $\le 2^{-\sqrt{\log(n)}}$.

\paragraph{\bf Statistical-Queries Lower bounds.}
Statistical queries (SQ) algorithms~\cite{Kearns93} is a class of learning algorithms whose interaction with $\cd$ is done only via {\em statistical queries}. Concretely, an SQ-algorithm can specify any function $Q:\{\pm 1\}^n\times\{\pm 1\}\to\{\pm 1\}$ and an error parameter $\lambda>0$, and receive a number $e$ satisfying $|E_{(x,y)\sim\cd}[Q(x,y)]-e|\le \lambda$. In this model, an algorithm is efficient if it makes polynomially many queries with error parameters satisfying $\frac{1}{\lambda}\le\poly(n)$ (besides that, the algorithm is not restricted). While this class is strictly smaller than the class of all algorithms, most known algorithms admit an SQ version. In addition, as opposed to general algorithms, unconditional lower bounds are known for SQ algorithms~\cite{blum1994weakly} for several learning problems. In particular, it is known that it is hard to agnostically learn halfspaces using SQ-algorithm (e.g. \cite{kalai2008agnostically}). We note that as with previously known lower bounds for general algorithms, these  results only rule out exact algorithms.

\paragraph{\bf Lower bounds on concrete algorithms.} A few results~\cite{Ben-DavidLoSreShr,
LongSe11, danielyLinSha13The} showing hardness of approximation results for several concrete families of algorithms (linear methods).

\paragraph{The methodology of \cite{daniely2013average}.}
% Computational vs Learning problems
In light of the great success of complexity theory in establishing hardness of approximation results for standard computational problems, having such dramatic gaps between upper and lower bounds is perhaps surprising. The reason for the discrepancy between learning problems and computational problems is the fact that it is unclear how to reduce $\mathbf{NP}$-hard problems to learning problems (see
\cite{ApplebaumBaXi08, daniely2013average}).
The main obstacle is the ability of a learning algorithm to return a hypothesis which does not
belong to the learnt class (in our case, halfspaces). Until recently, there was only a single
framework, due to Kearns and Valiant \cite{KearnsVa89}, to prove lower
bounds on learning problems. The framework of \cite{KearnsVa89} makes
it possible to show that certain cryptographic assumptions imply	
hardness of certain learning problems.  As indicated above, the lower
bounds established by this method are quite far from the
performance of best known algorithms.

% Our New method
In a recent paper \cite{daniely2013average} (see also
\cite{daniely2013more}) we, together with Linial and Shalev-Shwartz, developed a new framework to prove 
hardness of learning based on hardness on average of
CSP problems. Yet, in \cite{daniely2013average} we were not able
to use our technique to establish hardness results that are based on
a natural assumption on a well studied problem. Rather, we made a rather speculative hardness assumption, that is concerned with general $\csp$ problems, most of which were never studied explicitly. 
We recognized it as the main weakness of
our approach, and therefore the main direction for further research. About a year after, Allen, O'Donnell and Witmer~\cite{allenOdWi2015} refuted the assumption of \cite{daniely2013average}.
On the other hand we~\cite{danielySh2014} were able to overcome the use of our speculative assumption, and prove hardness of learning of DNF formulas (and other problems) based on a natural assumption on the complexity of refuting random $K$-SAT instances, in the spirit of Feige's
assumption~\cite{Feige02}. The current paper continues this line of work.

\subsection{Results}
\paragraph{Main result}
We say that a distribution $\cd$ on $\mathbb{Q}^n\times \{\pm 1\}$ is {\em $\eta$-almost realizable} if $\Err_{\half}(\cd)\le\eta$. An algorithm have {\em non-trivial} performance w.r.t. a certain family of distributions if for some $c>0$, its output hypothesis has error $\le \frac{1}{2}-\frac{1}{n^c}$ whenever the underlying distribution belongs to the family.
\begin{theorem}\label{thm:main}
\begin{itemize}
\item
Under assumption \ref{hyp:xor_weak}, for all $\eta>0$, there is no $\poly(n)$-time algorithm with non-trivial performance on $\eta$-almost-realizable distributions on $\{\pm 1\}^n\times\{\pm 1\}$.
\item
Under assumption \ref{hyp:xor_strong}, for all $\nu>0$, there is no $\poly(n)$-time algorithm with non-trivial performance on $2^{-\log^{1-\nu}(n)}$-almost-realizable distributions on $\{\pm 1\}^n\times\{\pm 1\}$.
\end{itemize}
\end{theorem}
These results imply in particular that under assumption \ref{hyp:xor_weak} there is no efficient learning algorithm with a constant approximation ratio, and under assumption \ref{hyp:xor_strong} the is no efficient learning algorithm with an approximation ratio of $2^{\log^{1-\nu}(n)}$. As mentioned above, this substantially improves on previously known results that only showed hardness of exact learning.

\paragraph{Extension to large margin learning}
Large margin learning is a popular variant of halfspace learning (e.g. \cite{ScholkopfSm02, Vapnik98}). Here, the learning algorithm faces an somewhat easier task, as it is not required to classify correctly examples that are very close to the separating hyperplane.
Our basic theorem can be extended to the large margin case.
Concretely, we say that a distribution $\cd$ on $\{\pm 1\}^n\times \{\pm 1\}$ is $\eta$-almost-realizable-with-margin if there is $w\in\mathbb{R}^d$ with $\sum_{i=1}^n|w_i|\le \poly(n)$ such that
$\Pr_{(x,y)\sim\cd}\left(y\cdot\inner{w,x}\le 1\right)\le \eta$.
Theorem \ref{thm:main} can be extended to show that no efficient algorithm can perform better than trivial even when the distribution is almost realized with margin. Concretely, we have the following theorem.
\begin{theorem}\label{thm:main_margin}
\begin{itemize}
\item
Under assumption \ref{hyp:xor_weak}, for all $\eta>0$, there is no $\poly(n)$-time algorithm with non-trivial performance on distributions on $\{\pm 1\}^n\times\{\pm 1\}$ that are $\eta$-almost-realizable-with-margin.
\item
Under assumption \ref{hyp:xor_strong}, for all $\nu>0$, there is no $\poly(n)$-time algorithm with non-trivial performance on 
distributions on $\{\pm 1\}^n\times\{\pm 1\}$ that are $2^{-\log^{1-\nu}(n)}$-almost-realizable-with-margin.
\end{itemize}
\end{theorem}

\paragraph{Statistical queries version}
Our proof technique can be adapted to show unconditional lower bounds for SQ-algorithms. Concretely, we have the following.

\begin{theorem}\label{thm:main_SQ}
There is no efficient SQ-algorithm with non-trivial performance on 
distributions on $\{\pm 1\}^n\times\{\pm 1\}$ that are $2^{-\log^{1-\nu}(n)}$-almost-realizable-with-margin.
\end{theorem}
Again, this substantially improves on previously known results that only showed hardness of exact learning.

\paragraph{Implications to hardness of approximation}
As explained in the previous section, the problem of proper learning is essentially equivalent to the problem of minimizing disagreements. As our results hold in particular for proper algorithms, we can conclude the following. 
\begin{theorem}\label{main:apx_hardness}
Under assumption \ref{hyp:xor_strong}, 
for every $\epsilon>0$ and $c>0$, no efficient algorithm can 
distinguish samples with error $\ge \frac{1}{2}-n^{-c}$ 
from samples with error $\le 2^{-\log^{1-\epsilon}(n)}$. 
\end{theorem}
We note that the conclusion of our theorem is stronger than the conclusions of the previously best known lower bounds for the problem (however, we rely on a stronger assumption). In particular, it implies that the problem is hard to approximate within a factor of $2^{\log^{1-\epsilon}(n)}$, implying the conclusion~\cite{arora1993hardness} of Arora et. al. It is also strengthen the conclusion~\cite{FeldmanGoKhPo06} of Feldman et. al., improving the completeness parameter from $2^{-\sqrt{\log(n)}}$ to $2^{-\log^{1-\epsilon}(n)}$ and the soundness parameter from $\frac{1}{2}-2^{-\sqrt{\log(n)}}$ to $\frac{1}{2}-n^{-c}$. We remark that our result holds even if we assume that the input examples are binary, and in the large margin settings.

\section{Main proof ideas}
We next elaborate on the main ideas of our main theorem. The full proof is deferred to the appendix.

\subsection{The methodology of \cite{daniely2013average}}
We first describe the methodology of \cite{daniely2013average} to prove hardness of learning. A {\em sample} is a collection $S=\{(x_1,y_1),\ldots,(x_m,y_m)\}\subset X\times \{\pm 1\}$. The {\em error} of $h:X\to \{\pm 1\}$ is $\Err_{h}(S)=\frac{|\{j:h(x_j)\ne y_j\}|}{m}$. The {\em error of $S$} w.r.t. a hypothesis class $\ch\subset \{\pm 1\}^X$ is $\Err_{\ch}(S)=\min_{h\in\ch}\Err_{h}(S)$.
The basic idea behind \cite{daniely2013average} is that if it is hard to distinguish a sample with small error form a sample that is very random in a certain sense, then it is hard to (even approximately) learn $\ch$.

For the sake of concreteness, we restrict to the problem of learning halfspaces over the boolean cube. Namely, we take $X=\{\pm 1\}^n$ and $\ch=\half=\{h_w\mid w\in \reals^n\}$.
We say that a sample is {\em strongly scattered\footnote{A weaker notion, called scattering, was used in \cite{daniely2013average}.}} if the labels (i.e., the $y_i$'s) are independent fair coins (in particular, they are independent from the $x_i$'s).
For $m=m(n)$ and $\eta=\eta(m)$, we denote by $\half_{m}^{\mathrm{s-scat},\eta}$ the problem of distinguishing a strongly-scattered sample from a sample with $\Err_{\half}(S)\le \eta$. Concretely, we say that the problem is easy if there exists an efficient randomized algorithm $\ca$ with the following properties. Its input is a sample $S=\{(x_1,y_1),\ldots,(x_m,y_m)\}\subset \{\pm 1\}^{n}\times \{\pm 1\}$, and its output satisfies:
\begin{itemize}
\item If $\Err_{\half}(S)\le \eta$, then
\[
\Pr_{\text{coins of }\ca}\left(\ca(S)=\text{``almost-realizable"}\right)\ge\frac{3}{4}
\]  
\item If $S$ is strongly scattered then, with probability $1-o_n(1)$ over the choice of the labels,
\[
\Pr_{\text{coins of }\ca}\left(\ca(S)=\text{``scattered"}\right)\ge \frac{3}{4}~.
\]  
\end{itemize}
\begin{theorem}\cite{daniely2013average} \label{lem_reduc_step_5}
If for every $a>0$ the problem $\half_{n^a}^{\mathrm{s-scat},\eta}$ is hard, then there is no efficient learning algorithm with non-trivial performance on $\eta$-almost realizable distributions.
\end{theorem}
To be self contained, and since Theorem \ref{lem_reduc_step_5} is not identical to \cite{daniely2013average}, we outline a proof.
\proof
Assume toward a contradiction that the efficient algorithm $\cl$ is guaranteed to return a hypothesis with error $\le \frac{1}{2}-\frac{1}{n^c}$ on $\eta$-almost realizable distributions. Let $M\left(n\right)$ be the maximal number of random bits used by $\cl$ when the examples lie in $\{\pm 1\}^n$. This includes both the bits describing the examples produced by the oracle and ``standard" random bits. Since $\cl$ is efficient, $M\left(n\right)< n^{c'}$ for some $c'>0$. Define
\[
q(n)=n^{2c'+2c}~.
\]
We will derive a contradiction by showing how 
$\cl$ can be used to solve $\half_{q(n)}^{\mathrm{s-scat},\eta}$. To this end, consider the algorithm $\ca$ defined below. On input $S=\{(x_1,y_1),\ldots,(x_m,y_m)\}\subset \{\pm 1\}^n\times\{\pm 1\}$,
\begin{enumerate}
\item Run $\cl$ on $S$, such that the examples' oracle generates examples by choosing random examples from $S$.
\item Let $h$ be the hypothesis that $\cl$ returns. If $\Err_S(h)\le \frac{1}{2}-\frac{1}{n^c}$, output $\text{``almost-realizable"}$. Otherwise, output $\text{``scattered"}$.
\end{enumerate}
Next, we show that $\ca$ solves $\half_{q(n)}^{\mathrm{s-scat},\eta}$. Indeed, if $\Err_{\half}(S)\le \eta$, then $\cl$ is guaranteed to return a hypothesis with $\Err_{S}(h)\le \frac{1}{2}-\frac{1}{n^c}$, and $\ca$ will output ``almost-realizable".
What if $S$ is strongly scattered? Let
$\cg\subset \{\pm 1\}^{\{\pm 1\}^n}$ be the collection of functions that $\cl$
might return. We note that $|\cg |\le 2^{n^{c'}}$, since  each
hypothesis in $\cg$ can be described by $n^{c'}$ bits. Namely, the
random bits that $\cl$ uses and the description of the examples sampled by the oracle. Now, since $\cd$ is
strongly-scattered, by Hoeffding's bound, the probability that $\Err_{S}(h)\le
\frac{1}{2}-\frac{1}{n^c}$ for a single $h:\{\pm 1\}^n\to \{\pm 1\}$ is $\le \exp\left(-2\frac{q(n)}{n^{2c}}\right)$. By the union bound, the probability that $\Err_{S}(h)\le
\frac{1}{2}-\frac{1}{n^c}$ for some $h\in \mathcal{G}$ is at most $|\cg|\exp\left(-2\frac{q(n)}{n^{2c}}\right)\le 2^{n^{c'}-\frac{q(n)}{n^{2c}}}= 2^{n^{c'}-n^{2c'}}$. It follows that the probability that $\ca$ responds
``almost-realizable" is $o(1)$.
\proofbox

\subsection{An overview}\label{sec:overview}
For the sake of simplicity, we will first explain how to prove a weaker version of Theorem \ref{thm:main}.
At the end of this section we will explain how to prove Theorem \ref{thm:main} in full.
\begin{theorem}\label{thm:simplified_main}
Suppose that for every $K> 4$ the problem $\csp_{n^{\frac{K}{4}}}^{\rand,1-\frac{1}{100}}(\xor_K)$ is hard. Then, there is no efficient algorithm with non-trivial performance on $\frac{2}{100}$-almost-realizable distributions on $\{-1,1,0\}^{n}\times\{\pm 1\}$.
\end{theorem}
The course of the proof is to reduce $\csp_{n^{\frac{K}{4}}}^{\rand,1-\frac{1}{100}}(\xor_K)$ to $\half_{n^{\frac{\sqrt{K}}{4\log(K)}}}^{\mathrm{s-scat},\frac{2}{100}}$. Given that reduction, and since $K$ can be arbitrarily large, Theorem \ref{thm:simplified_main} follows from Theorem \ref{lem_reduc_step_5}.

\subsubsection*{The $\xor$ problem as a learning problems}
The basic conceptual idea is to interpret the $K$-$\xor$ problem as a learning problems. Every $\psi\in \{\pm 1\}^n$ naturally defines $h_\psi:\cx_{n,K}\to \{\pm 1\}$, by mapping each $K$-tuple $C$ to the truth value of the corresponding constraint, given the assignment $\psi$. Namely, $h_{\psi}(C)=\xor(C(\psi))$.
Now, we can consider the hypothesis class $\ch_{K}=\{h_\psi\mid \psi\in\{\pm 1\}^n\}$.

The $K$-$\xor$ problem can be now formulated as follows. Given $J=\{C_1,\ldots,C_m\}\subset \cx_{n,K}$, find $h_\psi\in \ch_K$ with minimal error on the sample $S=\{(C_1,1),\ldots,(C_m,1)\}$. Now, the problem $\csp_{m}^{\rand,1-\frac{1}{100}}(\xor_K)$ is the problem of distinguishing between the case the $\Err_{\ch_K}(S) \le \frac{1}{100}$ and the case where the different $C_i$'s where chosen independently and uniformly from $\cx_{n,K}$.

The mapping $J\mapsto S$ is still not a reduction from $\csp_{m}^{\rand,1-\frac{1}{100}}(\xor_K)$ to the problem of distinguishing a strongly scattered sample from a sample with small error w.r.t. halfspaces. This is due to the following points:
\begin{itemize}
\item In the case that $J$ is random, $S$ is, in a sense, ``very random". Yet, it is not strongly-scattered.
\item We need to measure the error w.r.t. halfspaces rather than $\ch_{K}$.
\end{itemize}
Next, we explain how we address these two points.

\subsubsection*{Making the sample scattered}
Addressing the first point is relatively easy. Given a sample $(C_1,1),\ldots,(C_m,1)$, we can produce a new sample $(C'_1,y'_1),\ldots,(C'_m,y'_m)$ as follows: for every $i\in [m]$, w.p. $\frac{1}{2}$ we let $(C'_i,y'_i)=(C_i,1)$ and w.p. $\frac{1}{2}$ we let $(C'_i,y'_i)=(C'_i,-1)$, where $C'_i$ is obtained from $C_i$ by flipping the sign of the first literal. It is not hard to see that this reduction maps random instances to strongly scattered instances. Also, it is not hard to see that the error of every $h_\psi\in \ch_K$ does not change when moving from the original sample to the new sample. Therefore, samples with error 
$\le \frac{1}{100}$ are mapped to samples with error 
$\le \frac{1}{100}$.

Note that the reduction not only maps random instances to scattered instances, but in fact has a stronger property that will be useful for addressing the second point (Replacing $\ch_{K}$ by $\half$). Concretely, if the original instance is random, then the $(C'_i,y'_i)$'s are independent and uniform elements from $\cx_{n,K}\times\{\pm 1\}$. 

\subsubsection*{Replacing $\ch_{K}$ by $\half$}
To address the second point, we will show a reduction that maps a sample $S=\{(C_1,y_1),\ldots,(C_m,y_m)\}\subset \cx_{n,K}\times \{\pm 1\}$ to a sample $(x_1,y_1),\ldots,(x_m,y_m)\in \{-1,1,0\}^{n^{\sqrt{K}\log(K)}}\times \{\pm 1\}$. It will be convenient to give the reduction the option to output ``not-random" instead of producing a new sample.
For large enough $K$, if $m=n^{\frac{K}{4}}$, the reduction has the following properties:
\begin{itemize}
\item If the original sample is random (i.e., the examples $(C_i,y_i)$ where chosen independently and uniformly from $\cx_{n,K}\times \{\pm 1\}$), then w.h.p. the reduction will produce a new sample, and given that a new sample was indeed produced, it will be strongly scattered.
\item If the original sample has error $\le \frac{1}{100}$ w.r.t. $\ch_K$ then the reduction will either say that the original sample is not random, or it will produce a new sample with error $\le \frac{2}{100}$ w.r.t. halfspaces.
\end{itemize} 
Putting $n'=n^{\sqrt{K}\log(K)}$ and noting that $m=(n')^{\frac{\sqrt{K}}{4\log(K)}}$, it is not hard to see that such a reduction, together with the previous reduction (the ``scattering reduction"),  indeed reduces $\csp_{n^{\frac{K}{4}}}^{\rand,1-\frac{1}{100}}(\xor_K)$ to $\half_{n^{\frac{\sqrt{K}}{4\log(K)}}}^{\mathrm{s-scat},\frac{2}{100}}$.

The reduction will work as follows. It will first test that $J:=\{C_1,\ldots,C_m\}$ is {\em pseudo random}, in the sense that is satisfies certain (efficiently verifiable) properties (that will be specified later) that are possessed by random formulas w.h.p. 
If the test fails, the reduction will say that $S$ is not random. Otherwise, the reduction 
will produce the sample $\chi(S):=(\chi(C_1),y_1),\ldots,(\chi(C_m),y_m)\in \{-1,1,0\}^{n^{\sqrt{K}\log(K)}}\times \{\pm 1\}$, for a mapping $\chi:\cx_{n,K}\to \{-1, 1,0\}^{n^{\sqrt{K}\log(K)}}$ that we will specify later.

Next, we explain why this reduction have the desired properties. We start with the case that $S$ is sandom. In that case, the pseudo-randomness test will pass w.h.p. and the reduction will produce a new sample. Also, since the properties tested in this test do not depend on the labels, given that a new sample was indeed produced, the new sample is strongly scattered. 

We next deal with the case that $\Err_{\ch_K}(S)\le \frac{1}{100}$. We will show that in this case either the reduction will say that $S$ is not random, or the new sample will have error $\le \frac{2}{50}$ w.r.t. halfspaces. To this end we will finally specify $\chi$ and describe the list of pseudo-random properties.

It will be convenient to define $\chi$ as a composition $\chi=\rho\circ \pi$ where $\pi:\cx_{n,K}\to \{-1,1,0\}^{nK}$ 
and $\rho: \{-1,1,0\}^{nK}\to \{-1,1,0\}^{n^{\sqrt{K}\log(K)}}$. We first define $\pi$. The indicator vector of a literal is the vector in $\{0,-1,1\}^n$ whose all coordinates are zero except the coordinate corresponding to the literal, that is $1$ ($-1$) if the literal is un-negated (negated). We define $\pi(C)$ as a concatenation of $K$ vectors, where the $i$'s vector is the indicator vector of the $i$'th literal in $C$. As for $\rho$, we let $\rho(x)\in \{-1,1,0\}^{n^{\sqrt{K}\log(K)}}$ be a vector consisting of all products of the form $x_{i_1}\cdot\ldots\cdot x_{i_r}$ for $r\le \frac{1}{2}\sqrt{K}\log(K)$, and padded with zeros in the remaining coordinates. We note that for large enough $n$, the number of such products is $\le (nK+1)^{\frac{1}{2}\sqrt{K}\log(K)}\le n^{\sqrt{K}\log(K)}$.

Now, it is not hard to see that the error of $\chi(S)$ w.r.t. halfsapaces is exactly the error of $\pi(S)$ w.r.t. 
$\pol_d$ -- the hypothesis class consisting of threshold functions of polynomials of degree at most $d=\frac{1}{2}\sqrt{K}\log(K)$ .
Therefore, we will want to show that if $\Err_{\ch_K}(S)\le\frac{1}{100}$ then $\Err_{\pol_{d}}(\pi(S))\le\frac{2}{100}$.

Suppose that $\Err_{\ch_K}(S)\le\frac{1}{100}$, and let $\psi\in \{\pm 1\}^n$ such that $\Err_{h_\psi}(S)\le \frac{1}{100}$. 
To show that $\Err_{\pol_{d}}(\pi(S))\le\frac{2}{100}$ it is enough to construct a degree $\le d$ polynomial $p:\{-1,1,0\}^{nK}\to\mathbb R$ such that $\Pr_{j\sim [m]}\left(h_\psi(C_j)\ne\sign(p(\pi(C_j)))\right)\le\frac{1}{100}$.
Indeed, in that case, 
\begin{eqnarray*}
\Err_{\pol_d}(\pi(S)) &\le& \Pr_{j\sim [m]}\left(y_j\ne\sign(p(\pi(C_j)))\right)
\\
&\le& \Pr_{j\sim [m]}\left(y_j\ne h_\psi(C_j)\right)+\Pr_{j\sim [m]}\left(h_\psi(C_j)\ne \sign(p(\pi(C_j)))\right)
\\
&\le& \Err_{h_\psi}(S)+ \frac{1}{100}\le \frac{1}{100} + \frac{1}{100}=\frac{2}{100}
\end{eqnarray*}
We note that we will actually find $p$ with a stronger property - namely, that $\Pr_{j\sim [m]}\left(h_\psi(C_j)\neq p(\pi(C_j))\right)\le\frac{1}{100}$. This stronger property is not needed for proving the simplified version (Theorem \ref{thm:simplified_main}), but will be needed for proving Theorem \ref{thm:main}.

Unfortunately, it is not the case that we can always find such a polynomial. To overcome it, as explained above, the reduction will first check a set of pseudo-random properties. Concretely, for all small enough sets of literals, the reduction will check that the number of $K$-tuples that contain all the literals in the set is close to what is expected for a random sample. For example, it will check that the fraction of $K$-tuples containing the literal $x_7$ is approximately $\frac{K}{2n}$.

It therefore remains to show that for pseudo-random $S$ there is degree $\le d$ polynomial $p:\{-1,1,0\}^{nK}\to\mathbb R$ such that $\Pr_{j\sim [m]}\left(h_\psi(C_j)\ne p(\pi(C_j))\right)\le\frac{1}{100}$. To this end, consider the linear map $T_{\psi}:\mathbb R^{nK}\to \mathbb R^K$ that is defined as
\[
T_{\psi}\left(v_1|v_2|\ldots|v_K\right)=\left(\inner{v_1,\psi},\inner{v_2,\psi},\ldots,\inner{v_K,\psi}\right)
\]
We note that $\forall C\in\cx_{n,K},\;T_\psi(\pi(C))=C(\psi)$. We will consider polynomials of the form $p=p'\circ T_{\psi}$ where $p':\{\pm 1\}^K\to \reals$ is a degree $\le d$ polynomial. We note that for such $p$ we have 
\[
\Pr_{j\sim [m]}\left(h_\psi(C_j)\ne p(\pi(C_j))\right)=\Pr_{j\sim [m]}\left(\xor(C_j(\psi))\ne p'(C_j(\psi)))\right)
\]
It is therefore enough to find a degree $\le d$ polynomial $p'$ for which $\Pr_{z\sim \cd(\psi)}\left(\xor(z)=p'(z)\right)\ge 0.99$. Here, $\cd(\psi)$ is the distribution of the random variable $C_j(\psi)$ where $j\sim [m]$. Now, even though it is not possible to do that for general distributions, the fact that the sample is pseudo-random implies that $\cd(\psi)$ is close, in a certain sense, to the uniform distribution on $\{\pm 1\}^K$. 
Now, for uniform $z\in \{\pm 1\}^K$ we have that $\left|\sum_{i=1}^Kz_i\right|\le \frac{1}{2}\sqrt{K}\log(K)$ w.p. $1-o_K(1)$, and the same holds for $z\sim\cd(\psi)$. Therefore, since $\xor(z)$ is fully determined given $\sum_{i=1}^Kz_i$, and since $\sum_{i=1}^Kz_i$ takes $\le d$ values in the interval $[-d,d]$,
we can take $p'(z)=p''(\sum_{i=1}^Kz_i)$ where $p'':\reals\to\reals$ is a degree $\le \frac{1}{2}\sqrt{K}\log(K)$ polynomial that satisfies $p''(\sum_{i=1}^Kz_i)=\xor(z)$ whenever $\left|\sum_{i=1}^Kz_i\right|\le \sqrt{K}\log(K)$.

\subsubsection*{Proving Theorem \ref{thm:main} in full}
Theorem \ref{thm:simplified_main} differs from Theorem \ref{thm:main} in two aspects. The first is that in Theorem \ref{thm:main} the completeness parameter in the assumption can be arbitrarily close to $\frac{1}{2}$ (rather than $0.99$ in Theorem \ref{thm:simplified_main}) and at the same time, the conclusion holds for $\eta$-almost realizable distributions for arbitrarily small $\eta$ (rather than $\eta=0.02$ in Theorem \ref{thm:simplified_main}). The second aspect, that is much more minor, is that in Theorem \ref{thm:simplified_main} the hard distribution can be chosen to be supported in $\{-1,1,0\}^n\times\{\pm 1\}$, while Theorem \ref{thm:main} is slightly more restrictive and requires that the distribution will be supported in $\{\pm 1\}^n\times\{\pm 1\}$.

To address the first aspect, we will first reduce the random $\xor$ problem to the random majority-of-$q$-$\xor$s problem, and then we will follow a similar (but slightly more involved) argument as the one  described above. The reduction will work as follows. Given a $\xor$-formula with $m$ $\xor$-constraints, the reduction will produce a $\mxor$-formula with $\frac{m}{q}$ constraints, each of which is a majority of $q$ random $\xor$-constraints from the original formula. This reduction will amplify the completeness parameter toward $1$.

To address the second aspect, we note that halfspaces on $\{-1,1,0\}^n$ can be realized by halfspaces on $\{\pm 1\}^{2n}$ using the map $\Psi:\{-1,1,0\}^n\to \{\pm 1\}^{2n}$ that is defined as follows:
\[
\Psi(x)=(\Psi(x_1),\ldots,\Psi(x_n))~,
\]
where for $x\in \{0,-1,1\}$, $\Psi(x)=\begin{cases}(1,1) & x=1\\ (-1,-1) & x=-1\\ (-1,1) & x=0\end{cases}$. It is not hard to see that for every $w\in \mathbb R^{n}$ we have $h_w=h_{w'}\circ\Psi$ where $w'=\frac{1}{2}(w_1,w_1,\ldots,w_n,w_n)$. This observation shows that if there is no efficient algorithm with non-trivial performance on $\eta(n)$-almost-realizable distributions on $\{-1,1,0\}^n\times\{\pm 1\}$, then if there is no efficient algorithm with non-trivial performance on $\eta\left(\frac{n}{2}\right)$-almost-realizable distributions on $\{\pm 1\}^n\times\{\pm 1\}$.

\subsubsection*{A road map}
In section \ref{sec:pseudo_rand_formulas} we analyze elementary properties of random formulas, and define 
accordingly a notion of a pseudo-random formula. We also show that if $J=\{C_1,\ldots,C_m\}\subset \cx_{n,k}$ is a 
pseudo-random formula, then for $j\sim [m]$ and every assignment $\psi\in \{\pm 1\}^n$, the distribution of the 
random variable $C_j(\psi)\in \{\pm 1\}^K$ is close to the uniform distribution in a certain sense. In section \ref{sec:appr_by_pol} we show that for pseudo-random formulas, for every every assignment $\psi\in \{\pm 1\}^n$, the mapping $C_j\mapsto C_j(\psi)$ can be approximated by a low-degree polynomial. Finally, the full reduction is outlined in section \ref{sec:reduction}. In section \ref{sec:extension_proofs} we briefly explain how our argument can be extended to prove Theorems \ref{thm:main_margin} and \ref{thm:main_SQ}.

%\section{Open questions}\label{sec:future}

\paragraph{Acknowledgements:}
Amit Daniely is a recipient of the Google Europe Fellowship in Learning Theory, and this research was supported in part by this Google Fellowship.
The author thanks Uriel Feige, Vitaly Feldman, Nati Linial and Shai Shalev-Shwartz for valuable discussions.  Many thanks to Sarah R. Ellen, Ryan O'donnel and David Witmer for telling me about their paper~\cite{allenOdWi2015}, that inspired this work.

\bibliography{bib}

\appendix
\section{Proof of Theorem \ref{thm:main}}
\subsection{Pseudo-random formulas}\label{sec:pseudo_rand_formulas}
A {\em partial $K$-tuple} supported in $A\subset [K]$ is a mapping $C:\{\pm 1\}^n\to \{-1,1,*\}^K$ such that the output coordinates corresponding to $A$ are literals corresponding to $|A|$ different variables, and the remaining coordinates are the constant function $*$. The {\em size} of $C$ is $|A|$. We denote by $\cx_{n,K,A}$ the collection of partial $K$-tuples that are supported in $A$. We note that
\begin{equation}\label{eq:num_part_tuples}
|\cx_{n,K,A}|=(2n)(2n-2)\cdot\ldots\cdot (2n+2-2|A|)
\end{equation}
For $A\subset [K]$ we denote by $\Pi_A:\{\pm 1\}^K\to \{-1,1,*\}^K$ the function that maps all coordinates in $[K]\setminus A$ to $*$ and leaves the remaining coordinates unchanged. For $C\in\cx_{n,K}$ we denote by $C_A:\{\pm 1\}^n\to \{-1,1,*\}^K$ the partial $K$-tuple $\Pi_A\circ C$.
For a $K$-formula $J$, and a partial $K$-tuple $C$ we define the {\em frequency} of $C$ in $J$ as $\Fr_J(C)=\frac{|\{C'\in J : C'_A=C\}|}{m}$.
We note that the for random $J$, if $C$ is a partial $K$-tuple that is supported in a set $A$ of size $t$, then $\Fr_J(C)$ is an average of $m$ independent Bernulli variables with parameter $p_{n,t}:=\frac{1}{|\cx_{n,K,A}|}=\frac{1}{(2n)(2n-2)\cdot\ldots\cdot(2(n-t+1))}$. By Hoeffding's bound we have
\begin{lemma}\label{lem:unifom_fr_single}
Let $J=\{C_1,\ldots,C_m\}$ be a random formula. Then, for every partial $K$-tuple of size $t$ we have
$\Pr_{J}\left(\left|\Fr_J(C)-p_{n,t}\right|\ge \tau\right)\le 2\exp\left(-2m\tau^2\right)$
\end{lemma}
We say that $J$ is {\em $(t,\tau)$-pseudo-random} if $\left|\Fr_J(C)-p_{n,t'}\right|<\tau$ for every partial $K$-tuple $C$ of size $t'\le t$.
We say that $J$ is {\em $\tau$-pseudo-random} if it is $(K,\tau)$-pseudo-random.
By lemma \ref{lem:unifom_fr_single}, the fact that the number of partial $K$-tuples is
\[
1+\sum_{j=1}^K\binom{K}{j}(2n)(2n-2)\cdot\ldots\cdot (2n+2-2j)\le 2^{K}(2n)^K\le(2n)^{2K}~,
\]
and the union bound we have
\begin{lemma}\label{lem:unifom_fr_all}
Let $J=\{C_1,\ldots,C_m\}$ be a random formula. Then the probability that $J$ is not $\tau$-pseudo-random is at most
$(2n)^{2K}2\exp\left(-2m\tau^2\right)$
\end{lemma}
For a formula $J=\{C_1,\ldots,C_m\}\subset \cx_{n,K}$ and $\psi\in \{\pm 1\}^n$, we denote by $\cd(J,\psi)$ the distribution of the random variable $C_j(\psi)\in \{\pm 1\}^K$ where $j\sim\mathrm{Uni}([m])$.
A vector $z\in \{-1,1,*\}^K$ is {\em supported} in $A\subset [K]$ if $A=\{i\in [K]\mid z_i\ne *\}$. We say that a distribution $\cd$ on $\{\pm 1\}^K$ is {\em $(t,\mu)$-close to the uniform distribution} if for every $z\in \{-1,1,*\}^K$ that is supported in a set $A\in \binom{[k]}{\le t}$ we have that $\left|\Pr_{z'\sim \cd}\left(\Pi_A(z')=z\right)-2^{-|A|}\right|\le \mu$

\begin{lemma}\label{lem:psaudo_random_imply_uni}
If $J$ is $(t,\tau)$-pseudo-random then for every $\psi\in \{\pm 1\}^n$, $\cd(J,\psi)$ is $(t,n^t\tau)$-close to the uniform distribution.
\end{lemma}
\proof
Fix a vector $z\in \{-1,1,*\}^K$ supported in $A\in \binom{[K]}{\le t}$. We have
\begin{eqnarray*}
\Pr_{z'\sim \cd(J,\psi)}\left( \Pi_A(z')=z \right) &=& \Pr_{j\sim\Uni([m])}\left( (C_j)_A(\psi)=z \right)
\\
 &=& \frac{1}{m}|\{j\in[m] \mid (C_j)_A(\psi)=z\}|
\\
&=&\frac{1}{m}\sum_{C \in \cx_{n,K,A} \mid C(\psi)=z} |\{j\in[m] \mid (C_j)_A=C\}|
\\
&=&\sum_{C \in \cx_{n,K,A} \mid C(\psi)=z}\Fr_J(C)
\end{eqnarray*}
Denote $U=\{C \in \cx_{n,K,A} \mid C(\psi)=z\}$ and note that $|U|=n(n-1)\cdot\ldots\cdot (n-|A|+1)=\frac{|\cx_{n,K,A}|}{2^{|A|}}$.
By the $(t,\tau)$-pseudo-randomness of $J$ we have
\begin{eqnarray*}
\left|\Pr_{j\sim\Uni([m])}\left( (C_j)_A(\psi)=z \right)-2^{-|A|}\right| &=&\left|\sum_{C \in U}\Fr_J(C)-2^{-|A|}\right|
\\
&\le & \left|\sum_{C \in U}p_{n,|A|}-2^{-|A|}\right|+ |U|\tau
\\
&= & \left||U| p_{n,|A|}-2^{-|A|}\right|+\left|U \right|\tau = \left|U \right|\tau\le n^{|A|}\tau \le n^{t}\tau
\end{eqnarray*}
\proofbox

\subsection{Approximately realizing assignments by polynomials}\label{sec:appr_by_pol}
In this section we will show that when a formula $J=\{C_1,\ldots,C_m\}$ is pseudo-random then for every assignment $\psi\in \{\pm 1\}^n$ the mapping $C\mapsto \xor(C(\psi))$ can be approximately realized by a low-degree polynomial on $J$. Namely, there is a low degree polynomial $p:\cx_{n,K}\to \{\pm 1\}$ such that on most $K$-tuples $C\in J$, we have $p(C)=\xor(C(\psi))$.
To this end, we must represent $K$-tuples as vectors. 
The way we will do this is the following. Recall that the indicator vector of a literal is the vector in $\{0,-1,1\}^n$ whose all coordinates are zero except the coordinate corresponding to the literal, that is $1$ ($-1$) if the literal is un-negated (negated).
Also, we defined $\pi:\cx_{n,K}\to \{0,-1,1\}^{nK}$ such that $\pi(C)$ is a concatenation of $K$ vectors, where the $i$'s vector is the indicator vector of the $i$'th literal in $C$.

We will use the following version of Chernoff's bound from \cite{LinLur14}. 
\begin{lemma}\label{lem:chernoff}
Let $\cd$ be a distribution on $\{\pm 1\}^K$, $1\ge \beta>0$ and $\frac{1}{2} \le \alpha < \frac{1+\beta}{2} $. Assume that for every $z\in \{-1,1,*\}$ that is supported in a set $A$ of size $r=\lceil\beta K\rceil$ we have $\Pr_{z'\sim\cd}\left(\Pi_{A}(z')=z\right)\le \alpha^{r}$. Then
\[
\Pr_{z\sim\cd}\left(\left|\sum_{i=1}^Kz_i\right|\ge \beta K\right)\le 2\exp\left(-D\left(\frac{1+\beta}{2},\alpha\right)K\right)
\]
\end{lemma}

\begin{lemma}\label{lem:realizind_XOR_by_poly}
Let $\cd$ be a distribution on $\{\pm 1\}^K$ that is $(t,\mu)$-close to the uniform distribution, and let $d$ such that $\frac{K2^d\mu}{d} < d\le t$.
Then, there exists a degree $\le d$ polynomial $p:\{\pm 1\}^{K}\to \mathbb{R}$ such that
\[
\Pr_{z\sim \cd}\left(\xor(z)\ne p(z)\right)\le 2\exp\left(-D\left(\frac{1}{2}+\frac{d}{2K},\frac{1}{2}+\frac{2^{d-1}\mu}{d}\right)K\right)
\]
\end{lemma}
\proof
Let $\Lambda:\{\pm 1\}^K\to \reals$ be the linear mapping $\Lambda(z)=\sum_{i=1}^Kz_i$. We note that $\xor(z)$ is fully determined given $\Lambda(z)$. Namely there is $f:\reals\to \{\pm 1\}$ such that $\xor=f\circ \Lambda$. We note that the image of $\Lambda$ has at most $d+1$ values in the interval $[-d,d]$. Therefore, there is a degree $\le d$ polynomial $q:\reals\to\reals$ that coincides with $f$ on $\Lambda\left(\{\pm 1\}^K\right)\cap [-d,d]$. Consider now the degree $\le d$ polynomial $p=q\circ\Lambda$. We have that $p$ coincides with $\xor$ whenever $|\sum_{i=1}^Kz_i|\le d$. Hence, we have
\[
\Pr_{z\sim \cd}\left(\xor(z)\ne p(z)\right)\le \Pr_{z\sim\cd}\left(\left|\sum_{i=1}^Kz_i\right|\ge d\right)~.
\]
The lemma will now follow from Lemma \ref{lem:chernoff} when $\beta=\frac{d}{K}$ and $\alpha=\frac{1}{2}+\frac{2^{d-1}\mu}{d}$. It remains to show that the conditions of the Lemmas hold, namely, that $\alpha<\frac{1+\beta}{2}$ and that for every $z\in \{-1,1,*\}$ that is supported in a set $A$ of size $r=\lceil\beta K\rceil=d$ we have $\Pr_{z'\sim\cd}\left(\Pi_{A}(z')=z\right)\le \alpha^{d}$. The first conditions follows from the requirement that $\frac{K2^d\mu}{d} < d$. For the second condition, since $\cd$ is $(t,\mu)$-close to the uniform distribution and $d\le t$ we have
\begin{eqnarray*}
\alpha^d &=& \sum_{i=0}^d \binom{d}{i}\left(\frac{1}{2}\right)^{d-i}\left(\frac{2^{d-1}\mu}{d}\right)^{i}
\\
&\ge & \left(\frac{1}{2}\right)^{d}+d\left(\frac{1}{2}\right)^{d-1}\left(\frac{2^{d-1}\mu}{d}\right)
\\
&=& \left(\frac{1}{2}\right)^{d}+\mu
\\
&\ge & \Pr_{z'\sim\cd}\left(\Pi_{A}(z')=z\right)
\end{eqnarray*}
\proofbox

\begin{lemma}\label{lem:realizind_XOR_form_by_poly}
Let $J=\{C_1,\ldots,C_m\}\in \cx_{n,K}$ be $(t,\tau)$-pseudo-random, $\psi\in \{\pm 1\}^n$ and $d$ such that $\frac{K2^dn^d\tau}{d} < d\le t$.
Then, there exists a degree $\le d$ polynomial $p:\{0,-1,1\}^{nK}\to \mathbb{R}$ such that
\[
\Pr_{j\sim [m]}\left(\xor(C_j(\psi))\ne p(\pi(C_j))\right)\le 2\exp\left(-D\left(\frac{1}{2}+\frac{d}{2K},\frac{1}{2}+\frac{2^{d-1}n^d\tau}{d}\right)K\right)
\]
\end{lemma}
\proof 
Since $J$ is  $(t,\tau)$-pseudo-random and $d\le t$, it is also $(d,\tau)$-pseudo-random. By Lemma \ref{lem:psaudo_random_imply_uni} $\cd(J,\psi)$ is $(d,n^d\tau)$-close to the uniform distribution. Therefore, by Lemma \ref{lem:realizind_XOR_by_poly} there is degree $\le d$ polynomial $p'$ for which
\[
\Pr_{z\sim \cd(J,\psi)}\left(\xor(z)\ne p'(z)\right)\le 2\exp\left(-D\left(\frac{1}{2}+\frac{d}{2K},\frac{1}{2}+\frac{2^{d-1}n^d\tau}{d}\right)K\right)~.
\]
Now, let $T:\reals^{nK}\to \reals^K$ be the linear map
\[
T\left(v_1|v_2|\ldots|v_K\right)=\left(\inner{v_1,\psi},\inner{v_2,\psi},\ldots,\inner{v_K,\psi}\right)
\]
Note that for all $C\in \cx_{n,K}$, $T(\pi(C))= C(\psi)$. Consider the degree $\le d$ polynomial $p=p'\circ T$. We have
\begin{eqnarray*}
\Pr_{j\sim [m]}\left(\xor(C_j(\psi))\ne p(\pi(C_j))\right) &=& \Pr_{j\sim [m]}\left(\xor(C_j(\psi))\ne p'(T(\pi(C_j)))\right)
\\
&=& \Pr_{j\sim [m]}\left(\xor(C_j(\psi))\ne p'(C_j(\psi))\right)
\\
&=& \Pr_{z\sim \cd(J,\psi)}\left(\xor(z)\ne p'(z)\right)
\\
&\le& 2\exp\left(-D\left(\frac{1}{2}+\frac{d}{2K},\frac{1}{2}+\frac{2^{d-1}n^d\tau}{d}\right)K\right)
\end{eqnarray*}
\proofbox

\subsection{The reduction}\label{sec:reduction}

\subsubsection*{Step I: Amplifying the gap -- from $\xor$ to majority of $\xor$s}
For odd $q$, we define the predicate $\mxor_{q,K}:\{\pm 1\}^{qK}\to \{\pm 1\}$ by
\[
\mxor_{q,K}\left(z\right)=\maj\left(\xor(z_1,\ldots,z_k),\ldots,\xor(z_{(q-1)K+1},\ldots,z_{qK})\right)
\]
A {\em $(q,K)$-tuple} is an element in $\cx_{n,q,K}:=\left(\cx_{n,K}\right)^q$. For a $(q,K)$-tuple $C=(C^1,\ldots,C^q)$ and an assignment $\psi\in \{\pm 1\}^n$ we denote $C(\psi)=(C^1(\psi),\ldots,C^q(\psi))\in \{\pm 1\}^{qK}$.
A $(q,K)$-formula is a collection $J=\{C_1,\ldots,C_m\}$ of $(q,K)$-tuples.
An instance to the {\em $(q,K)$-$\mxor$ problem} is a {\em $(q,K)$-formula}, and the goal is to find an assignment $\psi\in \{\pm 1\}^n$ that maximizes
$\val_{\psi,\mxor}(J):=\frac{|\{j : \mxor_{q,K}(C_i(\psi))=1\}|}{m}$. We define the {\em value} of $J$ as $\val_\mxor(J):=\max_{\psi\in\{\pm 1\}^n}\val_{\psi,\mxor}(J)$. 
For $m=m(n)$, $(q,K)=(q(n),K(n))$ and $\frac{1}{2}>\eta= \eta (n) > 0$, we say that the problem $\csp^{\rand,1-\eta}_m(\mxor_{q,K})$ is easy, if there exists an efficient randomized algorithm, $\ca$ with the following properties. Its input is a $(q,K)$-formula $J$ with $n$ variables and $m$ constraints and its output satisfies:
\begin{itemize}
\item If $\val_{\mxor}(J)\ge 1-\eta$, then
\[
\Pr_{\text{coins of }\ca}\left(\ca(J)=\text{``non-random"}\right)\ge\frac{3}{4}
\]  
\item If $J$ is random\footnote{To be precise, the $(q,K)$-tuples are chosen uniformly, and independently from one another.} then, with probability $1-o_n(1)$ over the choice of $J$,
\[
\Pr_{\text{coins of }\ca}\left(\ca(J)=\text{``random"}\right)\ge \frac{3}{4}~.
\]  
\end{itemize}

\begin{lemma}\label{lem_reduc_step_0}
The problem $\csp_{m}^{\rand,\eta}(\xor_K)$ can be efficiently reduced to $\csp_{\lfloor\frac{m}{q}\rfloor}^{\rand,1-4\exp\left(-2\left(\eta-\frac{1}{2}\right)^2q\right)}(\mxor_{q,K})$.
\end{lemma}
We will use the following version of Chernoff's bound from \cite{LinLur14}. 
\begin{lemma}\label{lem:chernoff_2}
Let $\cd$ be a distribution on $\{0,1\}^q$ and $\eta<\frac{1}{2} $. Assume that for every $A\subset[q]$ we have $\Pr_{z\sim\cd}\left(\forall i\in A, z_i=1\right)\le \eta^{|A|}$. Then
\[
\Pr_{z\sim\cd}\left(\sum_{i=1}^qz_i \ge \frac{q}{2}\right)\le \exp\left(-D\left(\frac{1}{2},\eta \right)q\right) \le \exp\left(-2\left(\frac{1}{2}-\eta \right)^2q\right)
\]
\end{lemma}

\proof (of Lemma \ref{lem_reduc_step_0})
Given a $K$-formula $J=\{C_1,\ldots,C_m\}\in \cx_{n,K}$ we will produce a $(q,K)$-formula $J'$ as follows. We first randomly throw $m-m\lfloor\frac{m}{q}\rfloor$ of $J$'s $K$-tuples. Then, we randomly partition the remaining tuples into $\lfloor\frac{m}{q}\rfloor$ equally sized ordered bundles. For each such bundle $\{C_{j_1},\ldots,C_{j_q}\}$, we add to $J'$ the $(q,K)$-tuple $C=(C_{j_1},\ldots,C_{j_q})$. To see that the reduction works note that:
\begin{itemize}
\item
If $J$ is random then so is $J'$.
\item
Assume now that $\val_{\xor}(J)\ge 1-\eta$, and let $\psi\in \{\pm 1\}^n$ be an assignment with $\val_{\psi,\xor}(J)\ge 1-\eta$. Consider a single random bundle $\{C_{j_1},\ldots,C_{j_q}\}$. By Lemma \ref{lem:chernoff_2} we have that the probability that for most $K$-tuples in the bundle we have $\xor_K(C(\psi))=-1$ is $\le \exp\left(-2\left(\eta-\frac{1}{2}\right)^2q\right)$. Hence, $E\left[1-\val_{\mxor}(J')\right]\le \exp\left(-2\left(\eta-\frac{1}{2}\right)^2q\right)$. By Markov's inequality we have that $\val_{\mxor}(J')\ge 1-4\exp\left(-2\left(\eta-\frac{1}{2}\right)^2q\right)$ w.p. $\ge \frac{3}{4}$.
\end{itemize}

\proofbox

\subsubsection*{Step II: Making the sample scattered -- from $(\mxor)$ to $(\mxor,\neg \mxor)$}

A {\em labeled $(q,K)$-formula} is a collection $J=\{(C_1,y_1)\ldots,(C_m,y_m)\}\subset \cx_{n,q,K}$.
An instance to the {\em $(q,K)$-$(\mxor,\neg \mxor)$ problem} is a labeled $(q,K)$-formula, and the goal is to find an assignment $\psi\in \{\pm 1\}^n$ that maximizes
$\val_{\psi,\mxor}(J):=\frac{|\{j : \mxor_{q,K}(C_i(\psi))=y_i\}|}{m}$. We define the {\em value} of $J$ as $\val_\mxor(J):=\max_{\psi\in\{\pm 1\}^n}\val_{\psi,\mxor}(J)$. 
For $m=m(n)$, $(q,K)=(q(n),K(n))$ and $\frac{1}{2}>\eta= \eta (n) > 0$, we say that the problem $\csp^{\rand,1-\eta}_m(\mxor_{q,K},\neg \mxor_{q,K})$ is easy, if there exists an efficient randomized algorithm, $\ca$ with the following properties. Its input is a labeled $(q,K)$-formula $J$ with $n$ variables and $m$ constraints and its output satisfies:
\begin{itemize}
\item If $\val_{\mxor}(J)\ge 1-\eta$, then
\[
\Pr_{\text{coins of }\ca}\left(\ca(J)=\text{``non-random"}\right)\ge\frac{3}{4}
\]  
\item If $J$ is random\footnote{To be precise, the $(q,K)$-tuples and the labels are chosen uniformly, and independently from one another.} then, with probability $1-o_n(1)$ over the choice of $J$,
\[
\Pr_{\text{coins of }\ca}\left(\ca(J)=\text{``random"}\right)\ge \frac{3}{4}~.
\]  
\end{itemize}

\begin{lemma}\label{lem_reduc_step_1}
The problem $\csp_{m}^{\rand,1-\eta}(\mxor_{q,K})$ can be efficiently reduced to $\csp_{m}^{\rand,1-\eta}(\mxor_{q,K},\neg \mxor_{q,K})$.
\end{lemma}
\proof
Given an instance $J=\{C_1,\ldots,C_m\}$ to $\csp_{m}^{\rand,1-\eta}(\mxor_{q,K})$, we will produce an instance $J'$ to $\csp_{m}^{\rand,1-\eta}(\mxor_{q,K},\neg \mxor_{q,K})$ as follows.
For each $C_j$, w.p. $\frac{1}{2}$ we will add to $J'$ the 
pair $(C_j,1)$, and w.p. $\frac{1}{2}$ we will add the pair 
$(C'_j,-1)$ where $C'_j=(C'^1_j,\ldots,C'^q_j)$
is obtained 
from $C_j=(C_j^1,\ldots,C_j^q)$ by flipping, for each $C_j^i$, the sign of the first literal. It is not hard to see that 
if $J$ is random then so is $J'$. Also, for every $\psi\in \{\pm 1\}^n$, $\val_{\psi,\mxor}(J)=\val_{\psi,\mxor}(J')$, 
and therefore, if $\val_{\mxor}(J)\ge 1-\eta$ then $\val_{\mxor}(J')\ge 1-\eta$ as well.
\proofbox

\subsubsection*{Step III: Enforcing pseudo-randomness}
We say that a labeled $(q,K)$-formula $J$ is {\em $(t,\tau)$-pseudo-random} if the $K$-formula consisting of all the $K$-tuples that appear in $J$ is $(t,\tau)$-pseudo-random.
For $m=m(n)$, $(q,K)=(q(n),K(n))$, $(t,\tau)=(t(n),\tau(n))$ and $\frac{1}{2}>\eta= \eta (n) > 0$, we say that the problem $\csp_{m,(t,\tau)}^{\rand,1-\eta}(\mxor_{q,K},\neg\mxor_{q,K})$ is easy, if there exists an efficient randomized algorithm, $\ca$ with the following properties. Its input is a labeled $(q,K)$-formula $J$ with $n$ variables and $m$ constraints that is $(t,\tau)$-pseudo-random. Its output satisfies:
\begin{itemize}
\item If $\val_{\mxor}(J)\ge 1-\eta$, then
\[
\Pr_{\text{coins of }\ca}\left(\ca(J)=\text{``non-random"}\right)\ge\frac{3}{4}
\]  
\item If $J$ is random\footnote{To be precise, $J$ is chosen uniformly at random from all $(t,\tau)$-pseudo-random labeled $(q,K)$-formulas.} then, with probability $1-o_n(1)$ over the choice of $J$,
\[
\Pr_{\text{coins of }\ca}\left(\ca(J)=\text{``random"}\right)\ge \frac{3}{4}~.
\]  
\end{itemize}

\begin{lemma}\label{lem_reduc_step_2}
For $r=r(n) \ge 4$, the problem $\csp_{n^r}^{\rand,1-\eta}(\mxor_{q,K},\neg \mxor_{q,K})$ can be efficiently reduced to $\csp_{n^r,\left(r,n^{-\frac{r}{4}}\right)}^{\rand,1-\eta}(\mxor_{q,K},\neg\mxor_{q,K})$.
\end{lemma}
\proof
Given and instance $J$ to $\csp_{n^r}^{\rand,1-\eta}(\mxor_{q,K},\neg \mxor_{q,K})$ we will simply check weather it is $\left(r,n^{-\frac{r}{4}}\right)$-pseudo-random or not. If it not, we will say that $J$ is not random. Otherwise, we will leave it as is as an instance to $\csp_{n^r,\left(r,n^{-\frac{r}{4}}\right)}^{\rand,1-\eta}(\mxor_{q,K},\neg\mxor_{q,K})$. To see that this reduction works, note that
\begin{itemize}
\item
If $\val_{\mxor}(J)\ge \eta$, we will either say that it is not random or produce an $\left(r,n^{-\frac{r}{4}}\right)$-pseudo-random instance with value $\ge \eta$.
\item
If $J$ is random, by lemma \ref{lem:unifom_fr_all}, it is $\left(r,n^{-\frac{r}{4}}\right)$-pseudo-random with probability at least
\[
1-(2n)^K2\exp\left(-2n^{r}n^{-\frac{r}{2}}\right) = 
1-2\exp\left(K\log(2n)-2n^{r}n^{-\frac{r}{2}}\right)
\ge 1-o_n(1)~.
\]
Hence, w.p. $\ge 1-o_n(1)$ the reduction will produce an instance. Now, conditioning on this event, the produced formulas is a random labeled $(q,K)$-formula that is $\left(r,n^{-\frac{r}{4}}\right)$-pseudo-random.
\end{itemize}
\proofbox

\subsubsection*{Step IV: From $\mxor$ to polynomials}
Let $\pol_{u,d}$ be the hypothesis class of all functions $h:\{0,-1,1\}^u\to \{\pm \}$ that are thresholds of degree $\le d$ polynomials.
For $u=u(n),\;d=d(n),\;m=m(n)$ and $\eta=\eta(n)$ we consider the problem $\pol(u,d)_{m}^{\mathrm{s-scat},\eta}$ of distinguishing a strongly-scattered sample from a sample with $\Err_{\pol_{d}}(S)\le \eta$. Concretely, the input is a sample $S=\{(x_1,y_1),\ldots,(x_m,y_m)\}\subset \{-1,1,0\}^u\times \{\pm 1\}$, and we say that the problem is easy if there exists an efficient randomized algorithm, $\ca$ with the following properties. Its input is such a sample $S$, and its output satisfies:
\begin{itemize}
\item If $\Err_{\pol_{d}}(S)\le \eta$, then
\[
\Pr_{\text{coins of }\ca}\left(\ca(S)=\text{``almost-realizable"}\right)\ge\frac{3}{4}
\]  
\item If $S$ is strongly scattered then, with probability $1-o_n(1)$ over the choice of the labels,
\[
\Pr_{\text{coins of }\ca}\left(\ca(J)=\text{``scattered"}\right)\ge \frac{3}{4}~.
\]  
\end{itemize}

\begin{lemma}\label{lem_reduc_step_3}
For $d$ such that $\frac{K2^{d}n^{d}\tau}{d}<d\le t$, $\csp_{m,\left(t,\tau\right)}^{\rand,1-\eta}(\mxor_{q,K},\neg\mxor_{q,K})$ can efficiently reduced to $\pol(nqK,d)_{m}^{\mathrm{s-scat},\eta'}$ where
\[
\eta'=\eta+2q\exp\left(-D\left(\frac{1}{2}+\frac{d}{2K},\frac{1}{2}+\frac{2^{d-1}n^d\tau}{d}\right)K\right)
\]
\end{lemma}
\proof
Given a labeled $(q,K)$-formula $J=\{(C_1,y_1),\ldots,(C_m,y_m)\}\subset \cx_{n,q,K}\times \{\pm 1\}$ we will simply produce the sample $S=\{(\pi(C_1),y_1),\ldots,(\pi(C_m),y_m)\}\subset \{-1,1,0\}^{nqK}\times \{\pm 1\}$. Here, $\Pi:\cx_{n,q,K}\to \{-1,1,0\}^{nqK}$ is the mapping $\pi(C^1,\ldots,C^q)=(\pi(C^1),\ldots,\pi(C^q))$, where $\pi:\cx_{n,K}\to\{-1,1,0\}^{nK}$ is as defined in section \ref{sec:appr_by_pol}.

Clearly if $J$ is random then $S$ is scattered. It remains to show that if $\val_{\mxor}(J)\ge 1-\eta$ and $J$ is $(t,\tau)$-pseudo-random then there is a degree $\le d$ polynomial that errs on $\le \eta'$ fraction of the examples.

Indeed, let $\psi\in \{\pm 1\}^n$ be an assignment that satisfies $\ge 1-\eta$ fraction of $J$'th $(q,K)$-tuples. By lemma \ref{lem:realizind_XOR_form_by_poly}, there is a polynomial $p:\{-1,1,0\}^{nK}\to \mathbb{R}$ of degree $\le d$ that satisfies $p(\pi(C))=\xor(C(\psi))$ on  $1-2\exp\left(-D\left(\frac{1}{2}+\frac{d}{2K},\frac{1}{2}+\frac{2^{d-1}n^d\tau}{d}\right)K\right)$ fraction of $J$'s $K$-tuples (here, $J$'s $K$-tuples are the $K$-tuples that appear in one of $J$'s $(q,K)$-tuples).

Let $p':\{-1,1,0\}^{nqK}\to\reals$ be the degree $\le d$ polynomial $p'(x)=\sum_{j=1}^q p(z_{(j-1)K+1},\ldots,z_{jK})$.
It is not hard to check that $\sign\left(p'(\pi(C))\right)=\mxor(C(\psi))$ on $1-2q\exp\left(-D\left(\frac{1}{2}+\frac{d}{2K},\frac{1}{2}+\frac{2^{d-1}n^d\tau}{d}\right)K\right)$ fraction of $J$'s $(q,K)$-tuples. 

Therefore, we have that $\sign\left(p'(\pi(C_i))\right)\ne y_i$ for at most $\eta+2q\exp\left(-D\left(\frac{1}{2}+\frac{d}{2K},\frac{1}{2}+\frac{2^{d-1}n^d\tau}{d}\right)K\right)$ fraction of the samples in $S$. 
\proofbox

\subsubsection*{Step V: From polynomials to halfspaces}
Let $u=u(n),\;m=m(n)$ and $\eta=\eta(n)$. Similarly to $\pol(u,d)_{m}^{\mathrm{s-scat},\eta}$, we define $\half(u)_{m}^{\mathrm{s-scat},\eta}$ as the problem of distinguishing a strongly-scattered sample consisting of $m$ examples in $\{\pm 1\}^u\times \{\pm 1\}$, from a sample
with $\Err_{\half}(S)\le \eta$. 
\begin{lemma}\label{lem_reduc_step_4}
The problem $\pol(u,d)_{m}^{\mathrm{s-scat},\eta}$ can efficiently reduced to $\half\left(2(u+1)^d\right)_{m}^{\mathrm{s-scat},\eta}$
\end{lemma}
\proof
It will be convenient to decompose the reduction into two steps, where the second only deals with the issue of replacing $\{-1,1,0\}^{(u+1)^d}$ by $\{\pm 1\}^{2(u+1)^d}$.
Given a sample
\[
S=\{(x_1,y_1)\ldots,(x_m,y_m)\}\subset \{-1,1,0\}^u\times \{\pm 1\}
\]
the reduction will first produce the sample
\[
\rho(S)=\{(\rho(x_1),y_1)\ldots,(\rho(x_m),y_m)\}\subset \{-1,1,0\}^{(u+1)^d}\times \{\pm 1\}~,
\]
where $\rho:\{-1,1,0\}^u\to \{-1,1,0\}^{(u+1)^{d}}$ is defined as follows. We index the coordinates in $\{-1,1,0\}^{(u+1)^{d}}$ by the functions in $([u]\cup\{*\})^{[d]}$ and we let
\[
\forall f\in ([u]\cup\{*\})^{[d]},\;\;\rho_{f}(x)=\Pi_{j=1}^dx_{f(j)}~.
\]
(where $x_*:=1$). It is not hard to see that the collection of degree $\le d$ polynomial functions from $\{-1,1,0\}^u$ to $\reals$ equals to
\[
\{x\mapsto\inner{w,\rho(x)}\mid w\in \reals^{(u+1)^d}\}~.
\]
Hence, $\Err_{\pol_{u,d}}(S)=\Err_{\half}(\rho(S))$. 

In the second step the reduction will produce the sample
\[
\Psi(\rho(S))=\{(\Psi(\rho(x_1)),y_1)\ldots,(\Psi(\rho(x_m)),y_m)\}\subset \{\pm 1\}^{2(u+1)^d}\times \{\pm 1\}
\]
$\Psi:\{-1,1,0\}^{(u+1)^d}\to \{\pm 1\}^{2(u+1)^d}$ that is defined as follows:
\[
\Psi(x)=(\Psi(x_1),\ldots,\Psi(x_n))~,
\]
where for $x\in \{0,-1,1\}$, $\Psi(x)=\begin{cases}(1,1) & x=1\\ (-1,-1) & x=-1\\ (-1,1) & x=0\end{cases}$. It is not hard to see that for every $w\in \mathbb R^{(u+1)^d}$ we have $h_w=h_{w'}\circ\Psi$ where $w'=\frac{1}{2}(w_1,w_1,\ldots,w_{(u+1)^d},w_{(u+1)^d})$. Therefore we have
\[
\Err_{\half}(\Psi(\rho(S)))\le \Err_{\half}(\rho(S))=\Err_{\pol_{u,d}}(S)
\]
Also, it is clear that if $S$ is strongly scattered then so is $\Psi(\rho(S))$. To summarize, the reduction $S\mapsto \Psi(\rho(S))$ forms a reduction from $\pol(u,d)_{m}^{\mathrm{s-scat},\eta}$ to $\half\left(2(u+1)^d\right)_{m}^{\mathrm{s-scat},\eta}$
\proofbox

\subsubsection*{Connecting the dots}
We start with the hard problem $\csp_{n^{r}}^{\rand,\eta}(\xor_K)$. Using Lemma \ref{lem_reduc_step_0} and Lemma \ref{lem_reduc_step_1}, we reduce it to $\csp_{\lfloor\frac{n^{r}}{q}\rfloor}^{\rand,2^{-K}}(\mxor_{q,K},\neg\mxor_{q,K})$. Since $\eta$ is bounded away from $\frac{1}{2}$, this can be done with $q=C\cdot K$ for a constant $C=C(\eta)$. We can reduce farther to $\csp_{n^{r-1}}^{\rand,2^{-K}}(\mxor_{q,K},\neg\mxor_{q,K})$ (by simply throwing away random $\lfloor \frac{n^{r}}{q}\rfloor-n^{r-1}$ tuples from the input formula)

Now, we use Lemma \ref{lem_reduc_step_2} to reduce to $\csp_{n^{r-1},(r-1,n^{-\frac{r-1}{4}})}^{\rand,2^{-K}}(\mxor_{q,K},\neg\mxor_{q,K})$. Using Lemma \ref{lem_reduc_step_3}, for every $d$ such that $\frac{K2^dn^{d-r+1}}{d}<d\le r$ we can reduce to $\pol(nCK^2,d)_{n^{r-1}}^{\mathrm{s-scat},\eta'}$ for
\[
\eta'\le 2^{-K}+2CK\exp\left(-2\left(\frac{d}{2K}-\frac{2^{d-1}n^{d-r+1}}{d}\right)^2K\right)~.
\] 
We will choose $d$ such that $d=o(r)$ and $d=\omega\left(\sqrt{\log(K)K}\right)$ (the exact choice depend on the assumption we start with and will be specified later). It is not hard to check that for such a choice, it holds that for large enough $r$ and $K$, $\frac{K2^dn^{d-r+1}}{d}<d\le r-1$ and
$\eta'\le 2^{-C_1\frac{d^2}{K}}$ for a constant $C_1>0$.

Now, by Lemma \ref{lem_reduc_step_4} we can reduce farther to $\half(n^{4d})_{n^{r-1}}^{\mathrm{s-scat},\eta'}$. Putting $n'=n^{4d}$, we conclude that $\half(n')_{n'^{\frac{r-1}{4d}}}^{\mathrm{s-scat},\eta'}$. Since $d=o(r)$, this can be reduced farther (by simply randomly throwing examples from the input sample) to $\half(n')_{n'^{a}}^{\mathrm{s-scat},\eta'}$ for any constant $a>0$. By Theorem \ref{lem_reduc_step_5}, we conclude that there is no efficient learning algorithm that can return a hypothesis with non-trivial error on a distribution $\cd$ on $\{\pm 1\}^{n'}\times\{\pm 1\}$ that is $\eta'$-almost realizable by halfspaces.

For the choice of $d$ and the calculation of $\eta'$ in terms of $n'$, we split to two cases, according to the assumption we started with.

{\bf Case 1 (Assumption \ref{hyp:xor_weak}).} Here, $r=c\log(K)\sqrt{K}$ and $K$ is constant. We will choose $d=\log^{\frac{2}{3}}(K)\sqrt{K}$. We will have $\eta'\le 2^{-\Omega\left(\log^{\frac{4}{3}}(K)\right)}$. Since $K$ can be arbitrarily large, $\eta'$ can be arbitrarily small.

{\bf Case 2 (Assumption \ref{hyp:xor_strong}).} Here, $r=cK$ and $K=\log^s(n)$.
Here, we will choose $d=\frac{K}{\log\log{K}}$.
Note that
\begin{eqnarray*}
\eta' &\le& 2^{-C_3\frac{K}{(\log\log K)^2}}
\\
&=& 2^{-C_3\frac{\log^s(n)}{(\log\log K)^2}}
\\
&\le& 2^{-\log^{s-1}(n)}
\end{eqnarray*}
Now, $\log(n')=4d\log(n)=\frac{4}{\log\log K}\log^{s+1}(n)\le \log^{s+1}(n)$. Hence,
\[
\eta'\le 2^{-\log^{\frac{s-1}{s+1}}(n')}
\]
Since $s$ can be arbitrarily large, we can get $\eta'\le 2^{-\log^{1-\nu}(n')}$

\subsection{How to prove Theorems \ref{thm:main_margin} and \ref{thm:main_SQ}?}\label{sec:extension_proofs}
We next briefly explain how our argument can be extended to prove Theorems \ref{thm:main_margin} and \ref{thm:main_SQ}.

Theorem \ref{thm:main_margin} is proved analogously to Theorem \ref{thm:main}. The only difference is that we have to verify certain properties of the vector defining the halfspace that almost realizes the sample. Namely, we have to make sure that (i) the sum of its coefficients of is polynomial in the dimension and (ii) that whenever we guarantee that its prediction is correct, its inner product with the instance is $\ge 1$ in absolute value. These two facts can be straight forwardly verified, by carefully going over the proof.

Theorem \ref{thm:main_SQ} can also be proved analogously to Theorem \ref{thm:main}. The only difference is that we have to verify that (i) the problem we start with cannot be solved efficiently using statistical queries and that (ii) the reduction steps can be done using statistical queries. This  strategy can indeed be carried out, if one is using the result~\cite{FeldmanPeVe2015} of Feldman, Perkins and Vempala, that shows the SQ-hardness of the initial problem. However, a simpler strategy can be applied. Concretely, we can use our argument to reduce the uniform $K$-sparse-parity learning problem to the problems of learning halfspaces. In this parity problem, the learning algorithm is given an access to examples $(x,h(x))$ where $x\in \{\pm 1\}^n$ is uniformly distributed and $h$ computes the XOR of $K$ unknown variables. It is known~\cite{blum2003noise, blum1994weakly} that no SQ-algorithm for the problem  can return a classifier with error $\le \frac{1}{2}-n^{-o(K)}$ using $n^{o(K)}$ queries with error parameters $n^{-o(K)}$.
Lemma \ref{lem:realizind_XOR_by_poly} shows that for any XOR of $K$ variables, there is a polynomial threshold function of degree $d$ that agree with $h$ on all but $2\exp\left(-\frac{d^2}{2K}\right)$-fraction of the examples (w.r.t. the uniform distribution). As in the proof of Theorem \ref{thm:main}, this fact establishes a reduction to the problem of agnostically learning halfspaces. All is left to show is that this reduction can be implemented using statistical queries.

Indeed, the reduction is of the following form. It introduces a mapping $\Psi:\{\pm 1\}^n\to \{\pm 1\}^{n^r}$ and reduce the original learning problem to the problem of learning $(\Psi(x),h(x))$, where $x$ is sampled from the original distribution (the uniform distribution in our case). Now, given a statistical query $Q:\{\pm 1\}^{n^r}\times \{\pm 1\}\to \{\pm 1\}$, in order to evaluate $E_{x}\left[Q(\Psi(x),h(x))\right]$, we can simply query the oracle of the original problem with the function $\tilde{Q}(x,y):=Q(\Psi(x),y)$.

\end{document}